\documentclass[english,aps,pre,twocolumn]{revtex4-1}
\usepackage[T1]{fontenc}
\usepackage[latin9]{inputenc}
\usepackage{amsmath}
\usepackage{graphicx}
\usepackage{amssymb}

\makeatletter
\@ifundefined{textcolor}{}
{%
 \definecolor{BLACK}{gray}{0}
 \definecolor{WHITE}{gray}{1}
 \definecolor{RED}{rgb}{1,0,0}
 \definecolor{GREEN}{rgb}{0,1,0}
 \definecolor{BLUE}{rgb}{0,0,1}
 \definecolor{CYAN}{cmyk}{1,0,0,0}
 \definecolor{MAGENTA}{cmyk}{0,1,0,0}
 \definecolor{YELLOW}{cmyk}{0,0,1,0}
 }

\usepackage{pslatex}

\makeatother

\usepackage{babel}

\begin{document}

\title{Phase diagram of a model for a binary mixture of nematic molecules
on a Bethe lattice}

\author{E. do Carmo}
\email{educarmo@if.usp.br}
\author{A. P. Vieira}
\email{apvieira@if.usp.br}
\author{S. R. Salinas}
\email{ssalinas@if.usp.br}
\affiliation{Instituto de Física, Universidade de São Paulo, Caixa Postal 66318,
CEP 05314-970, São Paulo, Brazil}
\begin{abstract}
We investigate the phase diagram of a discrete version of the Maier-Saupe
model with the inclusion of additional degrees of freedom to mimic
a distribution of rodlike and disklike molecules. Solutions of this
problem on a Bethe lattice come from the analysis of the fixed points
of a set of nonlinear recursion relations. Besides the fixed points
associated with isotropic and uniaxial nematic structures, there is
also a fixed point associated with a biaxial nematic structure. Due
to the existence of large overlaps of the stability regions, we resorted
to a scheme to calculate the free energy of these structures deep
in the interior of a large Cayley tree. Both thermodynamic and dynamic-stability
analyses rule out the presence of a biaxial phase, in qualitative
agreement with previous mean-field results. 
\end{abstract}
\maketitle

\section{Introduction}

The recent characterization of biaxial nematic phases in thermotropic
liquid crystals \cite{key-1} has renewed the theoretical interest
in mechanisms leading to macroscopic biaxiality in such systems. In
general, two possibilities exist \cite{key-7}: (i) Either the systems
are composed of intrinsically biaxial mesogens or (ii) the systems
consist of a mixture of rodlike and disklike mesogens. Although unquestionable
experimental evidence of biaxiality has only surfaced for systems
of the first type, there have been hints from both experiments \cite{key-3}
and computer simulations \cite{key-4} that mixtures can escape the
curse of phase segregation and form stable biaxial nematic phases
under the appropriate conditions. In this paper we introduce a framework
which will allow future theoretical investigations of what such conditions
might be, starting from microscopic models and obtaining analytical
results which go beyond mean-field calculations. As a first application,
we show that a very simple choice of the interaction potential between
mesogens in a binary mixture leads to a biaxial state which is unstable
towards segregation, in agreement with previous mean-field results
\cite{Palffy}.

The inclusion of additional degrees of freedom in the Maier-Saupe
model has been used to mimic a mixture of rodlike and disklike molecules,
and to provide an explanation for the appearance of a biaxial nematic
structure \cite{Palffy,Henriques}. According to some investigations
for a mean-field Maier--Saupe model restricted to a discrete set of
orientations, which we call the Maier--Saupe--Zwanzig (MSZ) model,
the inclusion of a fixed distribution of shape variables leads to
a stable biaxial nematic phase. The biaxial region of the phase diagram
is separated by critical lines from two distinct uniaxial nematic
phases \cite{Henriques}, in qualitative contact with some experimental
phase diagrams for a lyotropic liquid mixture \cite{YuSaupe}. This
quenched polymorphism, however, which is generally used in solid-state
systems, may not be adequate for liquids and liquid crystalline systems,
with relatively short relaxation times, which may be better represented
by thermalized degrees of freedom. We then carried out a mean-field
investigation of the analogous MSZ model, with an annealed (or thermalized)
distribution of shape variables \cite{Docarmo}. At the mean-field
level, we have shown that a biaxial solution of this thermalized problem
is still present, but it becomes thermodynamically unstable, and therefore
physically unacceptable. 

In this article we report an analysis of a similar MSZ model on the
deep interior of a Cayley tree, also known as a Bethe lattice. The
main purpose of this investigation is the analysis of the global phase
diagrams under the effects of fluctuations and short-range correlations
neglected in the simple mean-field picture, but allowed for by the
structure of the Cayley tree. In Section 2, we define the Zwanzig
or discrete version of the Maier-Saupe model. The statistical problem
is formulated in terms of a set of nonlinear recursion relations,
whose fixed points correspond to solutions deep in the interior of
a large tree \cite{Baxter,Thompson}. This simple MSZ model displays
a first-order transition between a disordered and a uniaxial ordered
nematic phase, which is indicated by an overlap of the temperature
ranges of (dynamic) stability of the fixed points associated with
disordered and ordered structures. The location of this first-order
boundary comes from the application of an ingenious scheme, due to
Gujrati \cite{Gujrati}, which leads to the correct thermodynamic
free energy corresponding to each attractor, and avoids the well-known
pathologies associated with the surface of the Cayley tree \cite{Eggarter}.
In the infinite-coordination limit, we recover the well-known mean-field
results. In Section 3, we introduce the MSZ model for a thermalized
binary mixture of rodlike and disklike molecules. Again, the problem
is formulated as a set of recursion relations, whose fixed points
correspond to the physical solutions. Besides the attractors associated
with the disordered and two ordered uniaxial structures, we find an
additional fixed point associated with a biaxial nematic structure.
There is a large region of overlap of (dynamic) stability of the two
nematic uniaxial attractors. However, we find that the attractor associated
with biaxial nematic phase is always dynamicaly unstable. Also, by
using Gujrati's method, we show that the fixed point associated with
the biaxial structure corresponds to a larger free energy with respect
to the coexisting uniaxial attractors. This thermodynamic analysis
and the (dynamic) stability analysis rule out the physical presence
of an equilibrium biaxial phase in the Bethe lattice, in qualitative
agreement with the previous mean-field results.

\section{The MSZ model on a Cayley tree}

The Maier-Saupe model is described by the Hamiltonian\begin{equation}
\mathcal{H}=-A\sum_{(i,j)}\sum_{\mu,\nu=x,y,z}S_{i}^{\mu\nu}S_{i}^{\mu\nu},\label{msc1}\end{equation}
where $A$ is a positive parameter, the first sum is over neighboring
sites on a lattice of $N$ sites, and $S_{i}^{\mu\nu}$ is an element
of the traceless tensor \begin{equation}
S_{i}^{\mu\nu}=\frac{1}{2}(3n_{i}^{\mu}n_{i}^{\nu}-\delta_{\mu\nu}),\label{msc26}\end{equation}
where $n_{i}^{\mu}$ is the $\mu$-component of the unit vector associated
with the orientation of a molecular aggregate on site $i$. In the
Maier-Saupe-Zwanzig (MSZ) model we restrict the unit directors to
the Cartesian axes, $\vec{n}_{i}=\pm(1,0,0)$, $\pm(0,1,0)$, $\pm(0,0,1)$.
The canonical partition function is given by\begin{equation}
Z=\sum_{\{\vec{n}_{i}\}}\exp\left[\beta A\sum_{(i,j)}\sum_{\mu,\nu=x,y,z}S_{i}^{\mu\nu}S_{i}^{\mu\nu}\right],\end{equation}
where $\beta$ is the inverse of the temperature. Since $S_{i}^{\mu\nu}$
is invariant under the transformation $\vec{n}_{i}=-\vec{n}_{i}$,
the problem is reduced to the analysis of a three-state model, which
can be shown to lead to the same qualitative features of the continuous
Maier-Saupe model \cite{Deoliveirafigueiredo}.

We now formulate the MSZ model on a Cayley tree. In Figure \ref{cayley_1},
we represent some generations (or layers) of a (rooted) Cayley tree
of coordination $q=3$. Also, we illustrate one of the branches of
this tree, with a particular set of states of the nematic molecules.
Each of the $q$ branches of a Cayley tree with $n$ layers of sites
is built from $q-1$ branches with $n-1$ layers. Due to the well-known
thermodynamic pathologies associated with the surface of a Cayley
tree \cite{Eggarter}, our aim is to study only those sites deep in
the interior of the tree, which define the Bethe lattice. We thus
focus on the properties of the central site (the root) of a tree with
$n$ surrounding layers as representative of the properties of the
Bethe lattice obtained as $n\rightarrow\infty$. The partition function
of the MSZ model on such a tree is written as\begin{equation}
\mathcal{Z}_{n}=\mathcal{Z}_{n}^{x}+\mathcal{Z}_{n}^{y}+\mathcal{Z}_{n}^{z},\label{eq:fullpf}\end{equation}
 where $\mathcal{Z}_{n}^{\mu}$ ($\mu=x,y,z$) is the partial partition
function obtained when the molecule occupying the central site lies
along the $\mu$ axis.

\begin{center}
\begin{figure}[h]
\centering \includegraphics[width=0.9\columnwidth]{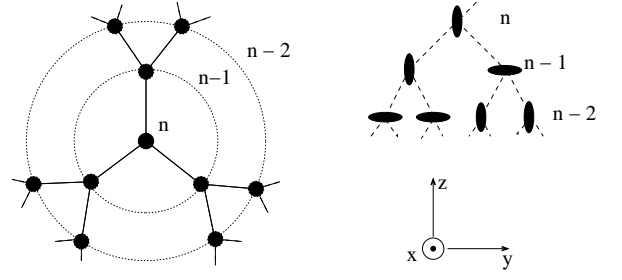} \caption{Some layers of a Cayley tree with coordination $q=3$. We also indicate
some microscopic states of the sites of one of the branches of this
tree.}

\label{cayley_1} 
\end{figure}

\par\end{center}

The connections with the usual nematic structures of the liquid-crystalline
systems are extracted from the tensor order parameter,\begin{equation}
Q^{\mu\nu}=\left\langle S_{0}^{\mu\nu}\right\rangle =\frac{3}{2}\left\langle n_{0}^{\mu}n_{0}^{\nu}\right\rangle -\frac{1}{2}\delta_{\mu\nu},\label{msc11}\end{equation}
 where $\left\langle \cdots\right\rangle $ indicates a thermal average,
and the subscript $0$ refers to the central site. Terms of the form
$\left\langle (n_{0}^{\mu})^{2}\right\rangle $ are related to the
density of nematic molecules (in the Bethe lattice) with the symmetry
axis along the $\mu$ direction (see Figure 1). We then adopt the
correspondence\begin{equation}
\left\langle (n_{0}^{\mu})^{2}\right\rangle \rightarrow\lim_{n\rightarrow\infty}\frac{\mathcal{Z}_{n}^{\mu}}{\mathcal{Z}_{n}}\label{param_a}\end{equation}
 according to which we have\[
Q_{n}^{\mu\mu}=-\frac{1}{2}+\frac{3}{2}\frac{\mathcal{Z}_{n}^{\mu}}{\mathcal{Z}_{n}},\]
 satisfying the traceless property\begin{equation}
Q_{n}^{xx}+Q_{n}^{yy}+Q_{n}^{zz}=0.\end{equation}
 It is convenient to introduce the variables\begin{equation}
S_{n}=Q_{n}^{zz}\end{equation}
 and\begin{equation}
\eta_{n}=Q_{n}^{yy}-Q_{n}^{xx},\end{equation}
 which are more adequate to distinguish between uniaxial and biaxial
nematic structures.

In order to calculate $S_{n}$ and $\eta_{n}$, we notice that $\mathcal{Z}_{n}^{\mu}$
depends on the sum over states of the molecules in each of the $q$
branches,\begin{equation}
\mathcal{Z}_{n}^{x}=\left[e^{\frac{3\beta A}{2}}\mathcal{Q}_{n}^{x}+e^{-\frac{3\beta A}{4}}\mathcal{Q}_{n}^{y}+e^{-\frac{3\beta A}{4}}\mathcal{Q}_{n}^{z}\right]^{q},\label{eq:zxq}\end{equation}
 \begin{equation}
\mathcal{Z}_{n}^{y}=\left[e^{-\frac{3\beta A}{4}}\mathcal{Q}_{n}^{x}+e^{\frac{3\beta A}{2}}\mathcal{Q}_{n}^{y}+e^{-\frac{3\beta A}{4}}\mathcal{Q}_{n}^{z}\right]^{q},\label{eq:zyq}\end{equation}
 and\begin{equation}
\mathcal{Z}_{n}^{z}=\left[e^{-\frac{3\beta A}{4}}\mathcal{Q}_{n}^{x}+e^{-\frac{3\beta A}{4}}\mathcal{Q}_{n}^{y}+e^{\frac{3\beta A}{2}}\mathcal{Q}_{n}^{z}\right]^{q},\label{eq:zzq}\end{equation}
where $\mathcal{Q}_{n}^{\mu}$ is the partial partition function of
a branch with $n$ layers obtained when the molecule in the single
site comprising the innermost layer lies along the $\mu$ axis. For
these partial partition functions of branches, it is straightforward
to write the recursion relations \begin{equation}
\mathcal{Q}_{n}^{x}=\left[e^{\frac{3\beta A}{2}}\mathcal{Q}_{n-1}^{x}+e^{-\frac{3\beta A}{4}}\mathcal{Q}_{n-1}^{y}+e^{-\frac{3\beta A}{4}}\mathcal{Q}_{n-1}^{z}\right]^{q-1},\label{msc4c}\end{equation}

\begin{equation}
\mathcal{Q}_{n}^{y}=\left[e^{-\frac{3\beta A}{4}}\mathcal{Q}_{n-1}^{x}+e^{\frac{3\beta A}{2}}\mathcal{Q}_{n-1}^{y}+e^{-\frac{3\beta A}{4}}\mathcal{Q}_{n-1}^{z}\right]^{q-1},\label{msc4b}\end{equation}
 and\begin{equation}
\mathcal{Q}_{n}^{z}=\left[e^{-\frac{3\beta A}{4}}\mathcal{Q}_{n-1}^{x}+e^{-\frac{3\beta A}{4}}\mathcal{Q}_{n-1}^{y}+e^{\frac{3\beta A}{2}}\mathcal{Q}_{n-1}^{z}\right]^{q-1}.\label{msc4a}\end{equation}

Defining the ratios\[
\rho_{n}^{x}=\frac{\mathcal{Q}_{n}^{x}}{\mathcal{Q}_{n}^{z}}\quad\mbox{and}\quad\rho_{n}^{y}=\frac{\mathcal{Q}_{n}^{y}}{\mathcal{Q}_{n}^{z}},\]
 Eqs. (\ref{msc4c}) to (\ref{msc4a}) lead to \begin{equation}
\rho_{n}^{x}=\left(\frac{r\rho_{n-1}^{x}+\rho_{n-1}^{y}+1}{\rho_{n-1}^{x}+\rho_{n-1}^{y}+r}\right)^{q-1}\equiv g_{x}\left(\rho_{n-1}^{x},\rho_{n-1}^{y}\right)\label{eq:rox}\end{equation}
 and\begin{equation}
\rho_{n}^{y}=\left(\frac{\rho_{n-1}^{x}+r\rho_{n-1}^{y}+1}{\rho_{n-1}^{x}+\rho_{n-1}^{y}+r}\right)^{q-1}\equiv g_{y}\left(\rho_{n-1}^{x},\rho_{n-1}^{y}\right),\label{eq:roy}\end{equation}
 with $r\equiv\exp\left(\frac{9}{4}\beta A\right)$. The connection
between the variables $\left(S_{n},\eta_{n}\right)$ and $\left(\rho_{n}^{x},\rho_{n}^{y}\right)$
is given by\begin{equation}
S_{n}=-\frac{1}{2}+\frac{3}{2}\frac{1}{\left[g_{x}\left(\rho_{n}^{x},\rho_{n}^{y}\right)\right]^{\frac{q}{q-1}}+\left[g_{y}\left(\rho_{n}^{x},\rho_{n}^{y}\right)\right]^{\frac{q}{q-1}}+1},\label{eq:sroxroy}\end{equation}
 \begin{equation}
\eta_{n}=\frac{3}{2}\frac{\left[g_{x}\left(\rho_{n}^{x},\rho_{n}^{y}\right)\right]^{\frac{q}{q-1}}-\left[g_{y}\left(\rho_{n}^{x},\rho_{n}^{y}\right)\right]^{\frac{q}{q-1}}}{\left[g_{x}\left(\rho_{n}^{x},\rho_{n}^{y}\right)\right]^{\frac{q}{q-1}}+\left[g_{y}\left(\rho_{n}^{x},\rho_{n}^{y}\right)\right]^{\frac{q}{q-1}}+1}.\label{eq:etaroxroy}\end{equation}

The problem is now reduced to the analysis of the set of two nonlinear
recursion relations (\ref{eq:rox}) and (\ref{eq:roy}). Physical
solutions on the Bethe lattice correspond to the (stable) fixed points
of the mapping problem, which are solutions of\[
\rho^{x}=g_{x}\left(\rho^{x},\rho^{y}\right)\quad\mbox{and}\quad\rho^{y}=g_{y}\left(\rho^{x},\rho^{y}\right).\]
 There is a trivial fixed point, $\rho^{x}=\rho^{y}=1$ ($S=\eta=0$),
corresponding to a disordered (isotropic) phase, and an ordered fixed
point, $\rho^{x}=\rho^{y}\neq1$ ($S\neq0$ and $\eta=0$), corresponding
to a uniaxial nematic structure. In this simple problem, it is easy
to see that there is no possibility of a biaxial nematic fixed point,
$S\neq0$ and $\eta\neq0$, which would require $\rho^{x}\neq\rho^{y}$.

The occurrence of a first-order transition is associated with the
existence of a range of temperatures in which both fixed points are
dynamically stable (in other words, both fixed points can be reached
from particular sets of initial conditions of the dynamic map \cite{Deoliveirasalinas}).
This is indeed the case of the nematic-isotropic transition, as can
be checked by a simple linear-stability analysis, i.e. by finding
the (degenerate) eigenvalues of the matrix\begin{equation}
\mathbf{M}=\left(\begin{array}{cc}
\frac{\partial g_{x}}{\partial\rho^{x}} & \frac{\partial g_{x}}{\partial\rho^{y}}\\
\frac{\partial g_{y}}{\partial\rho^{x}} & \frac{\partial g_{y}}{\partial\rho^{y}}\end{array}\right),\label{eq:matrizM}\end{equation}
 with the derivatives calculated at the fixed point of interest.

The trivial isotropic fixed point is stable for $\left\vert \Lambda_{\mathrm{iso}}\right\vert <1$,
where \begin{equation}
\Lambda_{\mathrm{iso}}=\left(q-1\right)\frac{1-\exp\left(-\frac{9}{4}\beta A\right)}{1+2\exp\left(-\frac{9}{4}\beta A\right)}.\end{equation}
 It is easy to find a numerical expression for the eigenvalue $\Lambda_{\mathrm{uni}}$,
which is associated with the limit of linear stability of the uniaxial
nematic fixed point, $\left\vert \Lambda_{\mathrm{uni}}\right\vert <1$.
In Fig. \ref{estab_1}, we draw graphs of $\Lambda_{\mathrm{iso}}$
and $\Lambda_{\mathrm{uni}}$ as a function of temperature, for a
typical value of the coordination $q$. The presence of a common range
of stability requires a detailed thermodynamic analysis to choose
the physical solution, corresponding to the smallest value of the
free energy. The problem is further delicate, due to the need to avoid
the pathologies produced by the surface sites of a Cayley tree.

\begin{center}
\begin{figure}[h]
\centering \includegraphics[width=0.9\columnwidth]{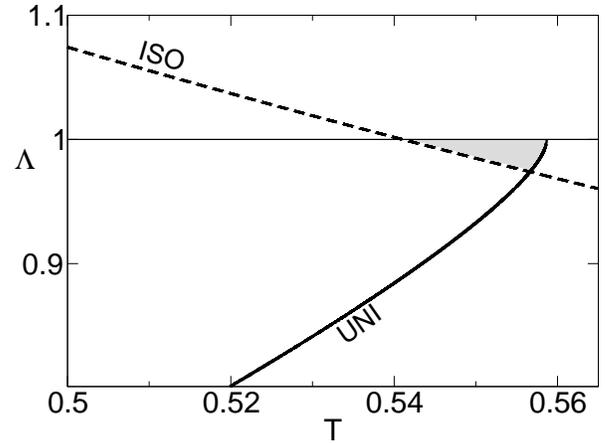} \caption{Temperature dependence of the eigenvalues associated with the analysis
of linear stability of the fixed points of the simple MSZ model on
a Cayley tree of coordination $q=3$. The grey region indicates the
joint range of stability ($|\Lambda_{\mathrm{iso}}|<1$ and $|\Lambda_{\mathrm{uni}}|<1$).}

\label{estab_1} 
\end{figure}

\par\end{center}

We now resort to a special technique \cite{Gujrati,Stilck} to find
the free energy associated with the bulk attractors of the recursion
relations. The idea of the method is to calculate that free energy
by cleverly subtracting the contribution from the surface sites. According
to an argument of Gujrati \cite{Gujrati}, we first write the free
energy of a Cayley tree with $M$ layers,\begin{eqnarray}
F_{M} & = & -\frac{1}{\beta}\ln\mathcal{Z}_{M}\nonumber \\
 & = & f_{0}^{(M)}+qf_{1}^{(M)}+q(q-1)f_{2}^{(M)}+\cdots\nonumber \\
 &  & +q(q-1)^{M-1}f_{M-1}^{(M)},\label{msc17}\end{eqnarray}
 where $\mathcal{Z}_{M}$ is the total partition function, defined
in Eq. (\ref{eq:fullpf}), and $f_{j}^{(M)}$ is the free energy per
site of molecules located at the $j$th layer of the tree, while the
coefficients of the $f_{j}^{(M)}$ are the number of sites in the
$j$th layer. We now rewrite this last equation as\[
F_{M}=f_{0}^{(M)}+q\sum_{j=1}^{M}\left(q-1\right)^{j-1}f_{j}^{(M)},\]
 and notice that for a tree with $M-1$ layers we can write\[
\left(q-1\right)F_{M-1}=\left(q-1\right)f_{0}^{(M-1)}+q\sum_{j=1}^{M-1}\left(q-1\right)^{j}f_{j}^{(M-1)}.\]
 (The factor $q-1$ enters due to the fact that a $M$-layer tree
has as many surfaces sites as $q-1$ trees with $M-1$ layers each.)
As $M\rightarrow\infty$, the free energies per site at the surface
of both trees, $f_{M}^{(M)}$ and $f_{M-1}^{(M-1)}$, should become
identical. In fact, we expect that $f_{j}^{(M)}-f_{j-1}^{(M-1)}\rightarrow0$
for all values of $j$. Thus,\begin{equation}
F_{M}-\left(q-1\right)F_{M-1}\rightarrow f_{0}^{(M)}+f_{0}^{(M-1)}.\label{eq:fmq-1fm-1}\end{equation}

\begin{figure}[h]
\centering \includegraphics[width=0.9\columnwidth]{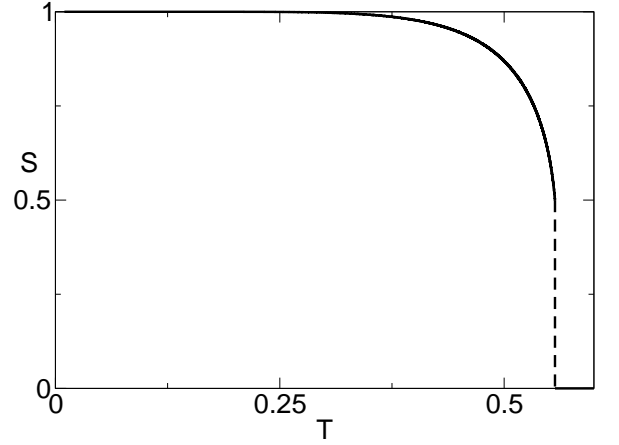} \caption{Graph of the order parameter $S$ versus temperature ($1/T=\beta Aq$)
for the MSZ model on a Bethe lattice of coordination $q=3$. The dashed
vertical line indicates a first-order transition. }
\label{sXt} 
\end{figure}

Since $f_{0}^{(M)}$ is the free energy associated with the central
site of a $M$-layer tree, and the properties of this central site,
for large $M$, are governed by the fixed points of Eqs. (\ref{eq:rox})
and (\ref{eq:roy}), we should have\[
\lim_{M\rightarrow\infty}f_{0}^{(M)}=\lim_{M\rightarrow\infty}f_{0}^{(M-1)}\equiv f_{b},\]
 the free energy per site $f_{b}$ thus representing all sites in
the Bethe lattice. Therefore, we conclude from Eqs. (\ref{msc17})
and (\ref{eq:fmq-1fm-1}) that\[
f_{b}=-\frac{1}{2\beta}\lim_{M\rightarrow\infty}\ln\frac{\mathcal{Z}_{M}}{\left(\mathcal{Z}_{M-1}\right)^{q-1}}.\]
 We then use Eqs. (\ref{eq:fullpf}) and (\ref{eq:zxq})-(\ref{msc4a})
in the limit $M\rightarrow\infty$ to write the free energy\begin{equation}
f_{b}\equiv f_{b}(q,A/T;S,\eta),\end{equation}
 where $S$ and $\eta$ are obtained from the fixed-point values of
$\rho^{x}$ and $\rho^{y}$ via Eqs. (\ref{eq:sroxroy}) and (\ref{eq:etaroxroy}),
and then determine which solution corresponds to thermodynamic equilibrium.
In Figure \ref{sXt}, we show a graph of the uniaxial order parameter,
$S$, in terms of temperature. The dashed line corresponds to a coexistence
of ordered ($S\neq0$) and disordered ($S=0$) solutions.

\section{Lattice model for a binary mixture of rods and disks}

Given a configuration $\left\{ \lambda_{i}\right\} $ of rodlike ($\lambda_{i}=+1$)
and disklike ($\lambda_{i}=-1$) molecular aggregates, the energy
of the MSZ model for a binary mixture is written as\begin{equation}
\mathcal{H}\left\{ \lambda_{i}\right\} =-A\sum_{(i,j)}\sum_{\mu\nu=x,y,z}\lambda_{i}S_{i}^{\mu\nu}\lambda_{j}S_{j}^{\mu\nu},\label{msc25}\end{equation}
 which leads to the canonical partition function\begin{equation}
Z_{B}={\displaystyle \sum_{\left\{ \lambda_{i}\right\} }}^{\prime}\sum_{\left\{ \overrightarrow{n}_{i}\right\} }\exp\left[\beta A\sum_{(i,j)}\sum_{\mu,\nu=x,y,z}\lambda_{i}\lambda_{j}S_{i}^{\mu\nu}S_{j}^{\mu\nu}\right].\end{equation}
 The sum over $\left\{ \lambda_{i}\right\} $ is restricted by the
fixed concentrations of the molecular types,\begin{equation}
N_{r}-N_{d}=\sum_{i=1}^{N}\lambda_{i},\quad N_{d}=N-N_{r},\end{equation}
 where $N_{r}$ ($N_{d}$) is the number of rodlike (disklike) molecules,
and $N$ is the total number of molecules. It is now convenient to
introduce a chemical potential and change to a grand-canonical ensemble,\begin{equation}
\Xi_{B}={\displaystyle \sum\limits _{N_{r}=0}^{N}}\exp\left(\beta\mu N_{r}\right)Z_{B}={\displaystyle \sum\limits _{\left\{ \lambda_{i}\right\} }}\sum_{\left\{ \overrightarrow{n}_{i}\right\} }\exp\left[-\beta\mathcal{H}_{\mathrm{eff}}\right],\end{equation}
 with unrestricted sums over the sets of variables, and the effective
Hamiltonian\begin{equation}
\mathcal{H}_{\mathrm{eff}}=-\frac{\mu}{2}\left(\sum_{i=1}^{N}\lambda_{i}+N\right)-A\sum_{(i,j)}\sum_{\mu,\nu=x,y,z}\lambda_{i}\lambda_{j}S_{i}^{\mu\nu}S_{j}^{\mu\nu}.\end{equation}

\begin{center}
\begin{figure}[h]
\centering \includegraphics[width=0.9\columnwidth]{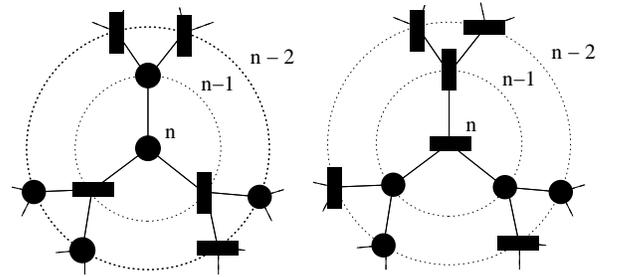} \caption{Two examples of configurations of rods and disks on the sites of a
Cayley tree of coordination $q=3$.}

\label{cayley_mix} 
\end{figure}

\par\end{center}

Along the lines of the treatment of Section 2 for the simple MSZ model
on a Cayley tree, we introduce some extra states to account for rod
and disk variables ($\lambda_{i}=\pm1$), and define a set of six
partial grand-canonical partition functions, $\Xi_{n}^{r,x}$, $\Xi_{n}^{r,y}$,
$\Xi_{n}^{r,z}$, $\Xi_{n}^{d,x}$, $\Xi_{n}^{d,y}$, $\Xi_{n}^{d,z}$,
associated with a tree of coordination $q$ with $n$ surrounding
layers (see Figure \ref{cayley_mix}). The notation is such that $\Xi_{n}^{r,z}$
represents the partial grand-canonical partition function of the tree
with the central site in a state with $\lambda_{i}=+1$ (rodlike)
and $\vec{n}_{i}$ along the $z$ direction, $\vec{n}_{i}=\pm(0,0,1)$.
In analogy to equations (\ref{eq:zxq}) to (\ref{eq:zzq}), it is
straightforward to write\begin{widetext}\begin{equation}
\Xi_{n}^{r,x}=e^{\beta\mu}\left[e^{\frac{3\beta A}{2}}\mathcal{Q}_{n}^{r,x}+e^{\frac{-3\beta A}{4}}(\mathcal{Q}_{n}^{r,y}+\mathcal{Q}_{n}^{r,z})+e^{\frac{-3\beta A}{2}}\mathcal{Q}_{n}^{d,x}+e^{\frac{3\beta A}{4}}(\mathcal{Q}_{n}^{d,y}+\mathcal{Q}_{n}^{d,z})\right]^{q},\label{eq:gczrx}\end{equation}
 \begin{equation}
\Xi_{n}^{r,y}=e^{\beta\mu}\left[e^{\frac{3\beta A}{2}}\mathcal{Q}_{n}^{r,y}+e^{-\frac{3\beta A}{4}}(\mathcal{Q}_{n}^{r,z}+\mathcal{Q}_{n}^{r,x})+e^{\frac{-3\beta A}{2}}\mathcal{Q}_{n}^{d,y}+e^{\frac{3\beta A}{4}}(\mathcal{Q}_{n}^{d,z}+\mathcal{Q}_{n}^{d,x})\right]^{q},\end{equation}
 \begin{equation}
\Xi_{n}^{r,z}=e^{\beta\mu}\left[e^{\frac{3\beta A}{2}}\mathcal{Q}_{n}^{r,z}+e^{-\frac{3\beta A}{4}}(\mathcal{Q}_{n}^{r,x}+\mathcal{Q}_{n}^{r,y})+e^{-\frac{3\beta A}{2}}\mathcal{Q}_{n}^{d,z}+e^{\frac{3\beta A}{4}}(\mathcal{Q}_{n}^{d,x}+\mathcal{Q}_{n}^{d,y})\right]^{q},\end{equation}
 \begin{equation}
\Xi_{n}^{d,x}=\left[e^{\frac{3\beta A}{2}}\mathcal{Q}_{n}^{d,x}+e^{\frac{-3\beta A}{4}}(\mathcal{Q}_{n}^{d,y}+\mathcal{Q}_{n}^{d,z})+e^{\frac{-3\beta A}{2}}\mathcal{Q}_{n}^{r,x}+e^{\frac{3\beta A}{4}}(\mathcal{Q}_{n}^{r,y}+\mathcal{Q}_{n}^{r,z})\right]^{q},\end{equation}
 \begin{equation}
\Xi_{n}^{d,y}=\left[e^{\frac{3\beta A}{2}}\mathcal{Q}_{n}^{d,y}+e^{\frac{-3\beta A}{4}}(\mathcal{Q}_{n}^{d,z}+\mathcal{Q}_{n}^{d,x})+e^{\frac{-3\beta A}{2}}\mathcal{Q}_{n}^{r,y})+e^{\frac{3\beta A}{4}}(\mathcal{Q}_{n}^{r,z}+\mathcal{Q}_{n}^{r,x})\right]^{q},\end{equation}
 \begin{equation}
\Xi_{n}^{d,z}=\left[e^{\frac{3\beta A}{2}}\mathcal{Q}_{n}^{d,z}+e^{\frac{-3\beta A}{4}}(\mathcal{Q}_{n}^{d,x}+\mathcal{Q}_{n}^{d,y})+e^{\frac{-3\beta A}{2}}\mathcal{Q}_{n}^{r,z}+e^{\frac{3\beta A}{4}}(\mathcal{Q}_{n}^{r,x}+\mathcal{Q}_{n}^{r,y})\right]^{q},\label{eq:gczdz}\end{equation}
\end{widetext}the quantities like $\mathcal{Q}_{n}^{d,x}$ now representing
partial grand-canonical partition functions of a branch with $n$
layers under the condition that the site in the innermost layer is
occupied by a disk whose symmetry axis lies along the $x$ axis. The
full partition function of the tree is written as\begin{equation}
\Xi_{n}=\Xi_{n}^{r,x}+\Xi_{n}^{r,y}+\Xi_{n}^{r,z}+\Xi_{n}^{d,x}+\Xi_{n}^{d,y}+\Xi_{n}^{d,z}.\end{equation}

We now use a similar parametrization as in Section 2. The tensor order
parameter is given by \begin{eqnarray}
Q^{\mu\nu} & = & \left\langle \frac{1}{N}\sum_{i}\lambda_{i}S_{i}^{\mu\nu}\right\rangle ,\nonumber \\
 & = & \frac{3}{2}\left\langle \frac{1}{N}\sum_{i}\lambda_{i}n_{i}^{\mu}n_{i}^{\nu}\right\rangle -\frac{1}{2}\delta_{\mu\nu}\left\langle \frac{1}{N}\sum_{i}\lambda_{i}\right\rangle \end{eqnarray}
where $\left\langle \cdots\right\rangle $ indicates a grand-canonical
thermal average. The factor $\left\langle \left(\sum_{i}\lambda_{i}\right)/N\right\rangle $
gives the difference between the concentrations (number fractions)
of rods and disks. It is then natural to introduce the correspondence\begin{equation}
\left\langle \frac{1}{N}\sum_{i}\lambda_{i}\right\rangle \rightarrow\lim_{n\rightarrow\infty}\frac{1}{\Xi_{n}}\sum_{\nu=x,y,z}\left[\Xi_{n}^{r,\nu}-\Xi_{n}^{d,\nu}\right].\label{msc38}\end{equation}
The ratio $\Xi_{n}^{r,z}/\Xi_{n}$ gives the concentration of rodlike
particles with the symmetry axis along the $z$ direction. Terms of
the form $\left(\sum_{i}\lambda_{i}(n_{i}^{\mu})^{2}\right)/N$ are
recognized as the difference between the concentrations of rods and
disks along the same $\mu$ direction, so that we have\begin{equation}
\left\langle \frac{1}{N}\sum_{i}\lambda_{i}(n_{i}^{\mu})^{2}\right\rangle \rightarrow\lim_{n\rightarrow\infty}\frac{\Xi_{n}^{r,\mu}-\Xi_{n}^{d,\mu}}{\Xi_{n}}.\label{paper2}\end{equation}
As in Section 2, it is interesting to work with the tensor order parameter
associated with the central site of a $n$-layer tree,\begin{equation}
Q_{n}^{\mu\mu}=-\frac{1}{2\Xi_{n}}\sum_{\nu}(\Xi_{n}^{r,\nu}-\Xi_{n}^{d,\nu})+\frac{3}{2}\frac{\Xi_{n}^{r,\mu}-\Xi_{n}^{d,\mu}}{\Xi_{n}},\label{msc41}\end{equation}
 such that\begin{equation}
\sum_{\mu}Q_{n}^{\mu\mu}=0.\end{equation}
 Again, it is convenient to introduce parameters $S_{n}$ and $\eta_{n}$
to characterize the distinct phases.

We can now write recursion relations for the branch partial partition
functions $\mathcal{Q}_{n}^{\alpha,\nu}$ {[}formally obtained from
Eqs. (\ref{eq:gczrx})-(\ref{eq:gczdz}) by taking $\Xi_{n}^{\alpha,\nu}\rightarrow\mathcal{Q}_{n}^{\alpha,\nu}$,
$\mathcal{Q}_{n}^{\alpha,\nu}\rightarrow\mathcal{Q}_{n-1}^{\alpha,\nu}$,
and $q\rightarrow q-1${]}, and rewrite them in terms of the set of
five ratios\[
\rho_{n}^{r,x}=\frac{\mathcal{Q}_{n}^{r,x}}{\mathcal{Q}_{n}^{r,z}},\quad\rho_{n}^{r,y}=\frac{\mathcal{Q}_{n}^{r,y}}{\mathcal{Q}_{n}^{r,z}},\]
\[
\rho_{n}^{d,x}=\frac{\mathcal{Q}_{n}^{d,x}}{\mathcal{Q}_{n}^{r,z}},\quad\rho_{n}^{d,y}=\frac{\mathcal{Q}_{n}^{d,y}}{\mathcal{Q}_{n}^{r,z}},\quad\rho_{n}^{d,z}=\frac{\mathcal{Q}_{n}^{d,z}}{\mathcal{Q}_{n}^{r,z}},\]
whose connection to the physical parameters $S_{n}$ and $\eta_{n}$
is easily determined from Eqs. (\ref{eq:gczrx})-(\ref{eq:gczdz}).
The resulting five-dimensional nonlinear mapping problem can be investigated
as in the previous section. The linear-stability analysis of the fixed
points is then reduced to studying the eigenvalues of a $5\times5$
matrix, analogous to that defined in Eq. (\ref{eq:matrizM}).

Depending on the values of temperature and chemical potential, there
appears a biaxial nematic fixed point ($S\neq0$ and $\eta\neq0$)
besides two distinct uniaxial nematic fixed points (with $S\neq0$
and $\eta=0$); of course, there is also an isotropic fixed point
($S=0$ and $\eta=0$). In the $\mu\times T$ plane (see Fig. \ref{est_mix1}),
there is a large low-temperature region of stability of both uniaxial
fixed points. However, the fixed point associated with the biaxial
nematic structure is dynamically unstable.

\begin{center}
\begin{figure}[h]
\centering \includegraphics[width=0.9\columnwidth]{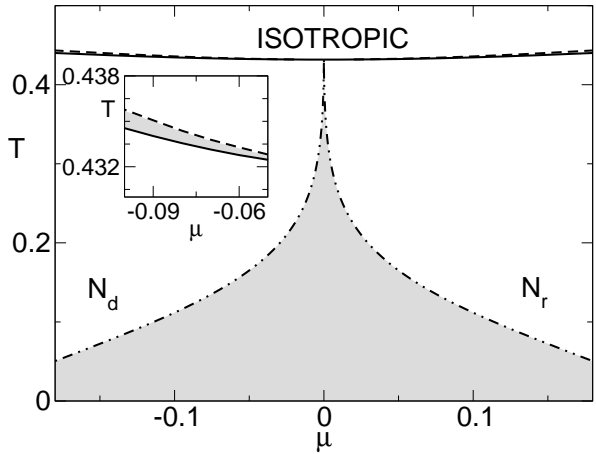} \caption{Stability lines in the $\mu\times T$ plane for a mixture of rodlike
and disklike molecules on a Cayley tree with coordination $q=3$.
Above the continuous line the isotropic phase is dynamically stable;
below the dashed line at least one of the nematic uniaxial phases
is dynamically stable; and between the dotted-dashed lines the two
nematic uniaxial phases are dynamically stable. The biaxial nematic
phase is dynamically unstable. }

\label{est_mix1} 
\end{figure}

\par\end{center}

The existence of common regions of stability of distinct fixed points
requires the analysis of the associated free energy in order to sort
out the physically acceptable phases and to locate the coexistence
(first-order) boundaries. Again, we use Gujrati's method to deal with
the subtleties of a Cayley tree. As in Section 2, the free energy
is written in terms of the function \begin{equation}
f_{b}^{\mathrm{mix}}\equiv f_{b}^{\mathrm{mix}}(q,A/T,\mu;\rho^{r,x},\rho^{r,y},\rho^{d,x},\rho^{d,y},\rho^{d,z}).\label{freemix}\end{equation}

\begin{figure}
\begin{centering}
\includegraphics[width=0.9\columnwidth]{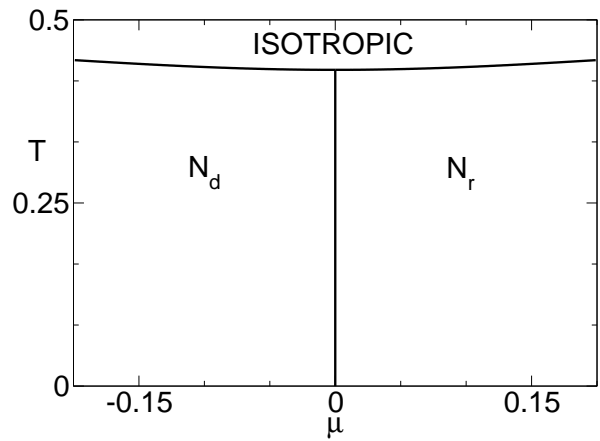}
\par\end{centering}

\caption{Phase diagram of the MSZ model for a mixture of rods and disks on
the Bethe lattice (of coordination $q=3$). We indicate two uniaxial
nematic ($N_{r}$ and $N_{d}$) and an isotropic phase, which are
bordered by first-order transition lines.}

\label{Flo:fig6}
\end{figure}

The phase diagram for a tree of a typical coordination, $q=3$, is
shown in Figure \ref{Flo:fig6}. At high temperatures, there is an
isotropic phase ($S=\eta=0$). At low temperatures, there are two
distinct uniaxial nematic phases, $N_{r}$, with $S>0$ and $\eta=0$,
with an excess of rods, and $N_{d}$, with $S<0$ and $\eta=0$, with
an excess of disklike particles. In this thermalized formulation of
the MSZ model, besides the biaxial nematic fixed point be dynamically
unstable, it is associated with larger values of the free energy as
compared to the uniaxial solutions, and cannot be thermodynamically
acceptable for all coordinations $q\geq3$. Of course, in the limit
of infinite coordination ($q\rightarrow\infty$, $A\rightarrow0$,
with $qA$ fixed), we recover all of the well-known results of the
mean-field calculations. Finally, in Figure \ref{estb_uni_iso_free}
we show the dynamic-stability lines of the isotropic and of the uniaxial
nematic phases. As expected, the first-order transition line lies
between the two stability lines.%

\begin{center}
\begin{figure}[h]
\centering \includegraphics[width=0.9\columnwidth]{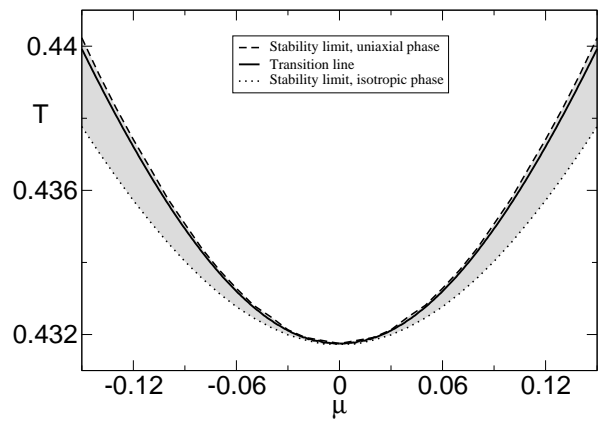} \caption{Phase diagram presenting the nematic-isotropic first-order transition
line and the two lines which indicate the dynamic-stability limit
of these two phases. As expected the transition line lies in the region
(shown in grey) where there is an overlap of dynamically stability
of the isotropic and the uniaxial nematic phases.}

\label{estb_uni_iso_free} 
\end{figure}

\par\end{center}

From the experimental point of view, it is interesting to draw the
corresponding temperature-concentration phase diagram. We note that
$c=-\partial f_{b}/\partial\mu$, and perform a Legendre transformation
to eliminate the chemical potential. In Figure \ref{txc_dia}, we
draw the characteristic tie lines between coexisting uniaxial nematic
phases. The regions of coexistence of isotropic and nematic phases
are too narrow to be clearly represented in this phase diagram.

\begin{center}
\begin{figure}[h]
\centering \includegraphics[width=0.9\columnwidth]{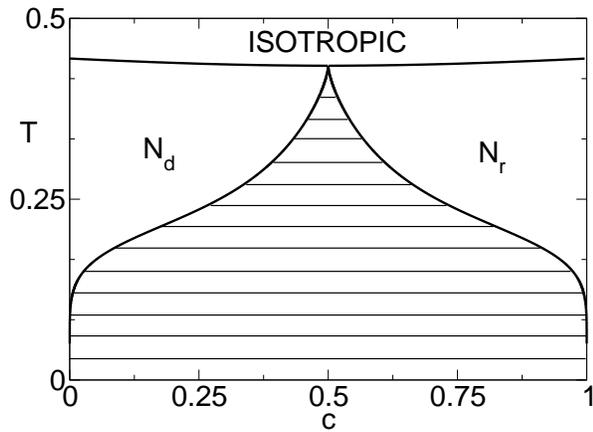} \caption{Temperature-concentration phase diagram of the MSZ model for a mixture
of rods and disks on a Bethe lattice of coordination $q=3$. The tie
lines indicate the coexistence of the uniaxial nematic phases, $N_{r}$
(rodlike molecules rich) and $N_{d}$ (disklike molecules rich). There
is no stable biaxial nematic phase.}

\label{txc_dia} 
\end{figure}

\par\end{center}

\section{Conclusions and perspectives}

We have formulated the problem of a Maier-Saupe-Zwanzig model on a
Cayley tree as a set of nonlinear discrete recursion relations, whose
attractors correspond to physically acceptable solutions on the Bethe
lattice (deep in the interior of the tree). Due to the presence of
first-order transitions, we have large regions of overlap of stability
of distinct attractors. We then resort to an ingenious scheme to obtain
the Bethe-lattice free energy associated with the coexisting attractors
and to determine the thermodynamically acceptable solutions.

We first considered the simple MSZ model on a Cayley tree. In this
problem, there are just two fixed points, corresponding to disordered
and uniaxial ordered structures, which are both (dynamically) stable
in a certain intermediate range of temperatures, and which is an indication
of occurrence of a first-order transition. We then use Gujrati's method
to locate the first-order boundary between the high temperature isotropic
and the low temperature uniaxial nematic phases. The MSZ model for
a binary mixture comes from the introduction of a new set of shape
variables and the definition of an effective Hamiltonian in the grand-canonical
ensemble. The problem is more involved, but we can perform a relatively
simple stability analysis of the fixed points, and draw a phase diagram
in terms of temperature and chemical potential. At low temperatures,
we find large regions of overlap of stability of fixed points associated
with uniaxial phases. However the biaxial structure is unstable both
from a dynamic-stability and a thermodynamic standpoint. The main
qualitative features of the phase diagrams on the Bethe lattice agree
with previous mean-field predictions \cite{Docarmo}.

The framework employed in the study of the binary mixture can be extended
in straightforward ways to deal with more complicated interaction
potentials, although of course requiring greater analytical effort.
An obvious extension is to consider a pair interaction described by\[
\sum_{\mu,\nu}\left[\lambda_{i}\lambda_{j}+b\left(\lambda_{i}+\lambda_{j}\right)+c\right]S_{i}^{\mu\nu}S_{j}^{\mu\nu},\]
which breaks the symmetry between rod-rod and disk-disk interactions.
It is also possible to introduce dilution and isotropic repulsive
forces to mimic the interaction potentials leading to stable biaxial
phases in recent computer simulations \cite{key-4}. We hope to report
on such extensions in future publications. Other extensions possibly
leading to stabilization of a biaxial phase in mixtures, such as the
introduction of polidispersity \cite{key-5,key-6}, can also be considered,
at least in principle, although the analytical work becomes quite
involved. 

%\vspace{1cm}
\begin{acknowledgments}
We acknowledge financial support from the Brazilian agencies CNPq
and FAPESP.\end{acknowledgments}


\begin{thebibliography}{18}
\bibitem{key-1}B. R. Acharya et al., Pramana \textbf{61}, 231 (2003);
B. R. Acharya, A. Primak, and S. Kumar, Phys. Rev. Lett. \textbf{92},
145506 (2004); L. A. Madsen, T. J. Dingemans, M. Nakata, and E. T. Samulski, Phys. Rev. Lett. \textbf{92},
145505 (2004); K. Merkel et al., Phys. Rev. Lett. \textbf{93}, 237801
(2004).

\bibitem{key-7}See e.g. R. Berardi et al., J. Phys.: Condens. Matter
\textbf{20}, 463101 (2008).

\bibitem{key-3}D. Apreutesei and G. H. Mehl, Chem. Commun. \textbf{2006},
609 (2006).

\bibitem{key-4}A. Cuetos, A. Galindo, and G. Jackson, Phys. Rev.
Lett. \textbf{101}, 237802 (2008).

\bibitem{Palffy}P. Palffy-Muhoray, J. R. de Bruyn, and D. A. Dunmur,
Mol. Cryst. Liq. Cryst. \textbf{127}, 301 (1985), and J. Chem. Phys.
\textbf{82}, 5294 (1985); S. R. Sharma, P. Palffy-Muhoray, B. Bergersen,
and D. A. Dunmur, Phys. Rev. A\textbf{ 32}, 3752 (1985).

\bibitem{Henriques}E. F. Henriques and V. B. Henriques, J. Chem.
Phys. \textbf{107}, 8036 (1997); E. F. Henriques, C. B. Passos, V.
B. Henriques, and L. Q. Amaral, Liquid Crystals \textbf{35}, 555 (2008).

\bibitem{YuSaupe}L. J. Yu and A. Saupe, Phys. Rev. Lett. \textbf{45},
1000 (1980); Y. Galerne and J. P. Marcerou, Phys. Rev. Lett. \textbf{51},
2109 (1983); A. A. de Melo-Filho, A. Laverde, and F. Y. Fujiwara,
Langmuir \textbf{19}, 1127 (2003).

\bibitem{Docarmo}E. do Carmo, D. B. Liarte, and S. R. Salinas, Phys.
Rev. E\textbf{ 81}, 062701 (2010).

\bibitem{Baxter}R. J. Baxter, \textit{Exactly solved models in statistical
mechanics}, Academic Press, New York, 1982.

\bibitem{Thompson}C. J. Thompson, J. Stat. Phys. z\textbf{ 27}, 441 (1982).

\bibitem{Gujrati}P. D. Gujrati, Phys. Rev. Lett. \textbf{74}, 809
(1995).

\bibitem{Eggarter}T. P. Eggarter, Phys. Rev. B \textbf{9}, 2989 (1974).

\bibitem{Deoliveirafigueiredo}M. J. de Oliveira, A. M. {Figueiredo
Neto}, Phys. Rev. A\textbf{ 34}, 3481 (1986).

\bibitem{DeGennes}P. G. de Gennes and J. Prost, \textit{The Physics
of Liquid Crystals}, Oxford University Press, Oxford, 1995.

\bibitem{Deoliveirasalinas}M. J. de Oliveira and S. R. Salinas, Rev.
Bras. Fis. \textbf{15}, 189 (1985).

\bibitem{Stilck}T. J. Oliveira, J. F. Stilck and P. Serra, Phys. Rev.
E\textbf{ 80}, 041804 (2009). 

\bibitem{key-5}Y. Mart\'{\i}nez-Rat\'on and J. A. Cuesta, Phys. Rev. Lett.
\textbf{89}, 185701 (2002).

\bibitem{key-6}L. Longa, G. Paj\k{a}k, and T. Wydro, Phys. Rev. E \textbf{76},
011703 (2007).
\end{thebibliography}
\end{document}